\documentclass[reprint,amsmath,amssymb,aps,superscriptaddress]{revtex4-2}
\usepackage{graphicx}
\usepackage{dcolumn}
\usepackage{bm}
\newcommand{\NTO}{Ni$_3$TeO$_6$}
\newcommand{\NISO}{Ni$_2$InSbO$_6$}

\begin{document}

\title{Spin Excitation in Coupled Honeycomb Lattice \NISO}
\author{Zheyuan Liu}
\affiliation{Institute for Solid State Physics, the University of Tokyo}
\author{Yusuke Araki}
\affiliation{Department of Advanced Materials Sciences, the University of Tokyo, Kashiwa 277-8561, Japan}
\author{Taka-hisa Arima}
\affiliation{Department of Advanced Materials Sciences, the University of Tokyo, Kashiwa 277-8561, Japan}
\author{Shinichi Itoh}
\affiliation{Institute of Materials Structure Science, High Energy Accelerator Research Organization, Ibaraki 305-0801, Japan}
\affiliation{Materials and Life Science Division, J-PARC Center, Tokai, Ibaraki 319-1195, Japan}
\author{Shinichiro Asai}
\affiliation{Institute for Solid State Physics, the University of Tokyo}
\author{Takatsugu Masuda}
\affiliation{Institute for Solid State Physics, the University of Tokyo}
\affiliation{Institute of Materials Structure Science, High Energy Accelerator Research Organization, Ibaraki 305-0801, Japan}
\affiliation{Trans-scale Quantum Science Institute, The University of Tokyo, Tokyo 113-0033, Japan}

\date{\today}

\begin{abstract}
We performed an inelastic neutron scattering experiment on a polycrystalline sample 
of a helimagnet \NISO \ to construct the spin Hamiltonian. 
Well-defined spin-wave excitation with a band energy of 20 meV was observed 
below $T_{N} = 76$ K. 
Using the linear spin-wave theory, the spectrum was reasonably reproduced 
with honeycomb spin layers coupled along the stacking axis (the $c$ axis). The proposed spin model reproduces the soliton lattice 
induced by a magnetic field applied perpendicular to the $c$ axis. 
\end{abstract}

\maketitle

\section{\label{sec:level1}Introduction}
In an insulating magnet with non-centrosymmetry, Dzyaloshinskii-Moriya (DM) interaction 
is activated through the spin-orbit coupling~\cite{PhysRev.120.91,DZYALOSHINSKY1958241}, 
favoring a non-collinear spin configuration, as opposed in a magnet which favors a collinear spin configuration owing to the symmetric exchange interaction. 
As a result of competition between the interactions, an incommensurate spin structure is often realized in the non-centrosymmetric magnet~\cite{dzyaloshinskii1964theory}. 
In the past decades, proper screw, conical, and cycloidal structures 
were experimentally observed in many compounds~\cite{PhysRev.151.414,doi:10.1143/JPSJ.49.545,Sosnowska_1982,McMorrow_1993,HAMACHER1997719,PhysRevB.48.6087,PhysRevB.59.11432,PhysRevLett.86.159}. Recently intriguing magnetic textures induced by a magnetic field were reported in these helimagnets. The spin soliton lattice, 
where the spin arrangement follows the solution of sine-Gordon equation, 
was observed when a magnetic field was applied parallel to the spin rotation plane~\cite{PhysRevLett.108.107202}. 
Control of the soliton lattice can be a significant technology 
in spintronics~\cite{doi:10.1126/science.1230155}. N{\'e}el-~\cite{bordacs2017equilibrium,PhysRevB.95.180410,kezsmarki2015neel,PhysRevLett.119.237201} and 
Bloch-type~\cite{yu2010real,doi:10.1126/science.1166767,doi:10.1126/science.1214143} 
skyrmions emerged in a magnetic field in polar and chiral systems, respectively. 
Stable magnetic skyrmions protected by non-trivial topology are promising candidates for novel magnetic memories 
devices~\cite{zhang2015skyrmion}. 
DM interaction is, thus, a key term for the emergent magnetic lattices in helimagnets.

\begin{figure}[b] 
\includegraphics[width=\linewidth]{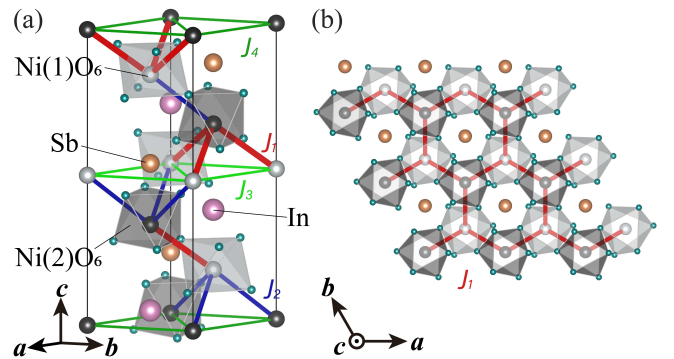}
\caption{\label{Fig_1}(a) Crystal structure of
\NISO\ and exchange
paths between Ni$^{2+}$ spins. Ni(1)O$_{6}$ octahedra (light gray), Ni(2)O$_{6}$ octahedra (dark gray), In$^{3+}$ ions, and Sb$^{5+}$ ions were displayed in a hexagonal unit cell. Oxygen ions around In$^{3+}$ and Sb$^{5+}$ ions are omitted. Exchange interactions, $J_{1}$ through $J_{4}$, are labeled in different colors. The thickness of each path indicates the magnitude of the exchange interaction. (b) Buckled honeycomb network formed by the dominant exchange interaction $J_{1}$ projected along the $c$ axis.} 
\label{fig1}
\end{figure}

In this study, we focus on a polar-chiral \NISO~\cite{doi:10.1021/cm304095s} where 
In$^{3+}$ and Sb$^{5+}$ ions are substituted for Ni$^{2+}$ and Te$^{6+}$ ions in the parent compound \NTO~\cite{oh2014non,doi:10.1021/acs.chemmater.7b01567,_ivkovi__2010,PhysRevLett.115.137201,PhysRevLett.117.147402} which is an ordered derivative of corundum. 
The space group is $R3$ with lattice parameters 
$a=5.2158$ \AA $^{-1}$ and $c=14.0139$ \AA $^{-1}$ in hexagonal notation, as shown in Fig.~\ref{Fig_1}(a). 
The ground state of \NISO\ was reported to be 
a proper-screw-type structure with the propagation vector of 
$\bm{k}=0.029\bm{b^{*}}$. 
The transition temperature was $T_{N}$ = 76 K. 
Recent study reported two-steps transitions; the compound first turns into a commensurate layered antiferromagnetic structure, 
and successively changes to the proper-screw-type structure 
which is generated by chirality-induced DM interaction \cite{PhysRevB.102.054409}. 
Ringlike magnetic scattering reported in several DM 
helimagnets~\cite{kezsmarki2015neel,PhysRevLett.119.237201,PhysRevB.95.144433,PhysRevB.97.020401} 
was observed by neutron diffraction experiments, 
suggesting that the direction of the propagation vector is, in fact, isotropic in the $ab$-plane~\cite{PhysRevB.102.054409}. 
In addition, the direction was controlled by in-plane magnetic field. 
Combination of magnetization, electric polarization, and dielectric constant measurements revealed 
an enriched phase diagram including helical, soliton lattice, canted antiferromagnetic (CAF) and 
${\bm q}$-flop phases~\cite{PhysRevB.102.054409}, 
offering an excellent environment to study the relationship of incommensurate magnetic structures and DM interaction. 

In conventional helimagnets with a fixed modulation axis, the soliton lattice is induced in the field applied in the 
spin rotation plane as a result of 
the competition between symmetric exchange and DM interaction responsible for the helical structure and Zeeman 
energy~\cite{PhysRevLett.78.4857,PhysRevLett.86.1885,PhysRevLett.108.107202,doi:10.7566/JPSJ.85.112001}. 
In contrasts in chiral polar helimagnet \NISO , the soliton lattice was proposed to 
be induced when a magnetic field was applied perpendicular to the spin rotation plane 
by assuming additional DM interactions with the staggered vector component along the polar $c$-axis~\cite{PhysRevB.102.054409}. 
However, the neighboring exchange interactions $J_1$ and $J_2$ in Fig.~\ref{fig1} (a) were 
proposed to be equivalent by Raman spectroscopy~\cite{PhysRevB.100.144417}, 
leading to uniform DM interaction instead of staggered one along the polar axis. 
To verify the scenario of the soliton lattice,
identification of precise spin Hamiltonian by measuring 
the spin excitation in large momentum - energy space is required.

Here we performed inelastic neutron scattering (INS) experiments on polycrystalline sample of \NISO\ at zero 
magnetic field. 
Well-defined spectra were observed and successfully analyzed by linear spin-wave theory. 
The estimated exchange constants, $J_1$ and $J_2$, were different by a factor of 6, 
supporting the scenario of 
the soliton lattice induced by the staggered DM interaction. 
In addition, chirality-induced DM interaction was discussed by combination of the present 
spin model and previously reported propagation vector. 
The estimated critical field applied parallel to the $c$ axis was consistent with 
the previous study~\cite{PhysRevB.102.054409}. 

\section{Experimental details}

Polycrystalline sample of \NISO\ with a mass of 21 g was synthesized by solid-state reaction method. 
The polycrystalline sample wrapped by aluminum foil were sealed in an Al cell. 
A Gifford-McMahon type cryostat was used to control the temperature down to 10 K. 
An inelastic neutron scattering experiment was performed by using 
High Resolution Chopper (HRC) spectrometer~\cite{ITOH201190} 
installed at BL-12 in MLF, J-PARC. 
The frequency of the T0 chopper was 50 Hz. 
The frequency of the Fermi chopper was 200 Hz, and the incident energies, $E_i$s, 
of 12.5, 15.3, 30.5, 61.2, and 102 meV were used for measurements at $T = 10$ K. 
An additional $E_i$ of 61.2 meV was used for the measurements at $T$= 10 K, 35 K, 65 K, 100 K, 300 K. 
The data reduction was performed by HANA software \cite{Kawana_2018}. 

\section{Experimental results}
\begin{figure}[htbp]
\includegraphics[width=\linewidth]{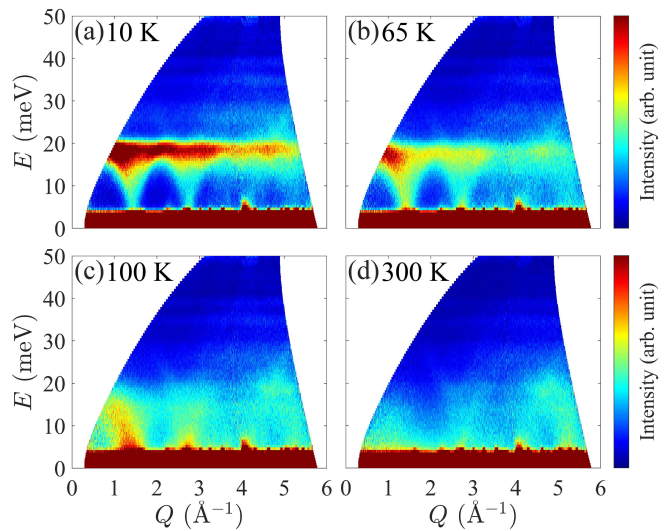}
\caption{\label{Fig_2}Inelastic neutron scattering spectra with the incident energy $E_{i} = 61.2\ $ meV measured at
(a)10 K, (b)65 K, (c)100 K, and (d)300 K.} 
\end{figure}
Temperature dependence of inelastic neutron scattering spectrum with $E_i$ = 61.2 meV 
is shown in Figs.~\ref{Fig_2}(a)-\ref{Fig_2}(d). 
$E$ and $Q$ denote the energy transfer and the momentum transfer, respectively. 
Well-defined spin-wave excitations with a band energy of 20 meV are
observed at 10 K and 65 K which are below $T_{N}$. 
The intensity at 65 K is weaker because the sub-lattice moment is reduced near $T_{N}$. 
The spectrum at 100 K is smeared, nevertheless the remnant feature of spin-wave excitation is 
observed due to short-range spin correlation, which is 
characteristic of a low-dimensional spin system. 
The spectrum at 300 K, which is well above $T_C$, is featureless for $Q \lesssim 2$ \AA .
The observed excitations for $Q \gtrsim 2$ \AA\ are due to phonons. 
The detailed structure of the spectrum with $E_i = 30.5$ meV at 10 K is shown in Fig.~\ref{Fig_3}(a). 
Spin-wave excitation with no anisotropy gap is observed in the 
energy range of $E \gtrsim 2$ meV. 
The anisotropy gap is not observed 
in the range of $E \gtrsim 0.6$ meV in the spectrum with $E_{i} = 12.5$ meV as well (not shown) . 

\section{Simulation}
\begin{figure*}
\includegraphics[width=\linewidth]{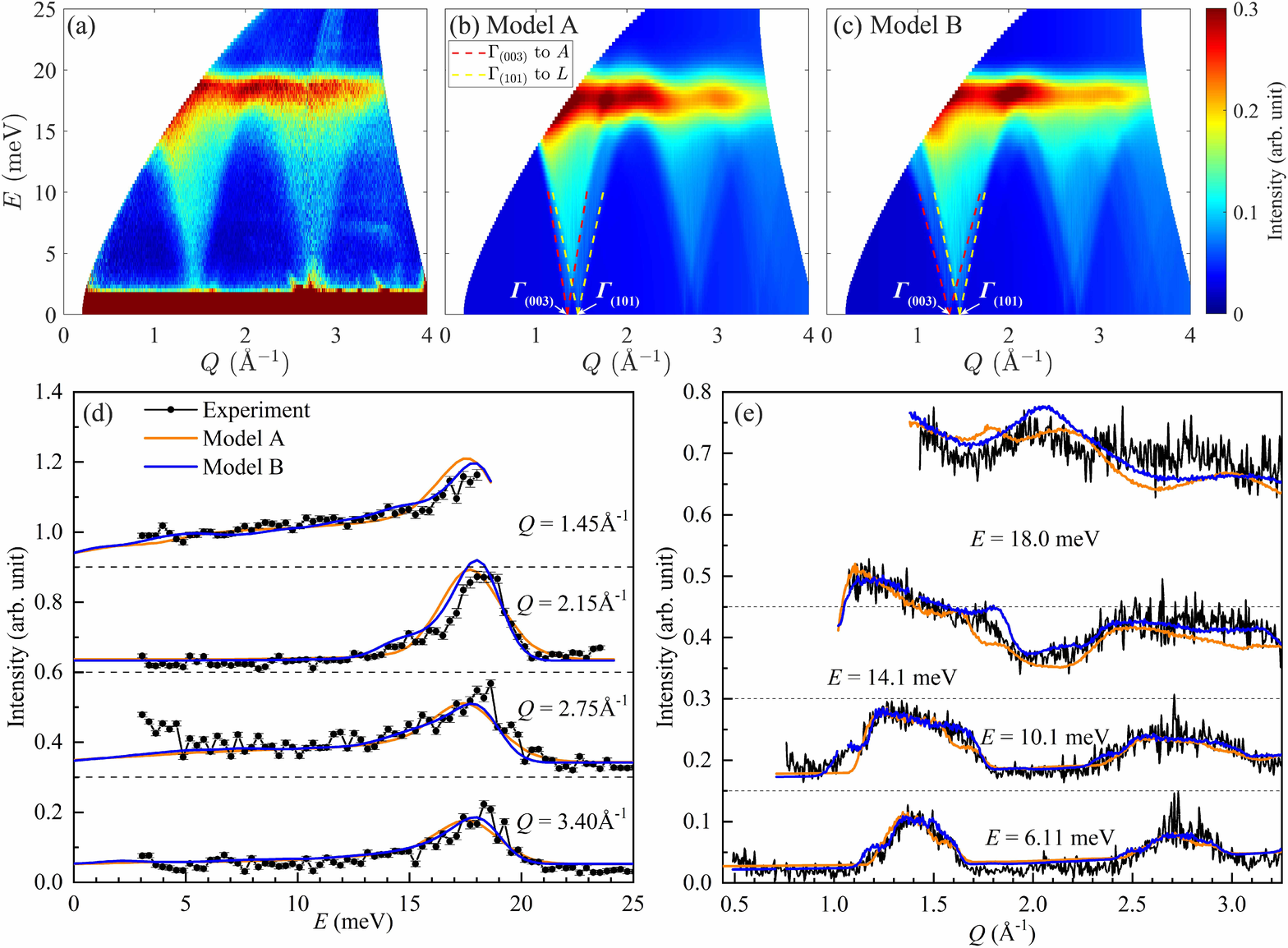}
\caption{\label{Fig_3}(a) INS spectrum with $E_{i}$=30.5 meV at $T$=10 K. (b), (c) Calculated INS spectra of polycrystalline sample using the best fit parameters for model A in (b) and model B in (c). Modes of spin wave excitation from $\Gamma _{(003)}$ to $A$ and from $\Gamma _{(101)}$ to $L$ points are described by red and yellow dash lines, respectively. The incoherent elastic scattering below 3 meV is excluded from the fitting area. (d),(e) 1D cuts along the energy transfer in (d) and those along the momentum transfer in (e), where orange and blue lines indicate the simulations using model A and model B, respectively. } 
\end{figure*}

Simulation of spin-wave excitation for
\NISO\ was performed based on 
linear spin-wave theory (LSWT) using SpinW package~\cite{toth2015linear}. 
Analytic approximation was adopted for the magnetic form factor of Ni$^{2+}$ ions~\cite{Dianoux2003NeutronDB}. In this section we identify the main part of the spin Hamiltonian which dominates the observed spin spectra. In our simulation we used a Heisenberg spin model without DM interaction terms 
and a collinear antiferromagnetic spin structure was assumed 
as the ground state, 
since the incommensurability of the spin structure is as small as 0.04 \AA $^{-1}$, 
which is also hard to be probed by the present experiment. 
DM interaction crucial for the phase diagram will be discussed by a combination of 
the present INS experiment and the previous studies on neutron diffraction and bulk property measurements 
in the forthcoming section. 

The Heisenberg spin Hamiltonian is represented as
\begin{equation}
{\mathcal H} = \sum_{\left\langle i,j\right\rangle }{J(\bm{r_i-r_j})\bm{S_{i}}\cdot \bm{S_{j}}}, 
\label{Eq_1}
\end{equation}
where $\bm{r}_{i}$ is the position of $i$th Ni$^{2+}$ ion and $\bm{S}_{i}$ is the spin operator at the $i$th Ni$^{2+}$ ion. 
The sum is taken over pairs of spins. 
In the crystal structure, two kinds of inequivalent Ni$^{2+}$ ions
denoted by Ni(1) and Ni(2) are stacked along the $c$-axis, three of each ion are contained in the hexagonal unit cell, as shown in Fig.~\ref{Fig_1}(a). 
The first and second neighbor exchange interactions occur via the Ni-O-Ni path, 
whereas the third and fourth neighbor 
exchange interactions occur via the Ni-O-O-Ni path; all bond angles are 
obtuse, suggesting antiferromagnetic super-exchange interactions for all the cases. The first, second, third and forth neighbor interactions are denoted by $J_{1}$, $J_{2}$, $J_{3}$ and $J_{4}$, shown in Fig.~\ref{Fig_1}(a). The labels for the spin model used here are the same as those in Ref.~\cite{PhysRevB.100.144417}. 

\begin{table}
\caption{\label{Tab_1}
Atomic distances and bond angles for exchange paths in \NISO . The exchange paths $J_{3}$ and $J_{4}$ go through two intervening oxygen atoms}
\begin{ruledtabular}
\begin{tabular}{ccccc}
&$J_{1}$&$J_{2}$&$J_{3}$&$J_{4}$\\
\hline
Ni$-$Ni (\AA)&3.747&3.876&5.216&5.216\\
$\angle$Ni$-$O$-$Ni($^{\circ}$)&128.34&134.95&-&-\\
\end{tabular}
\end{ruledtabular}
\end{table}

We tried a couple of models to fit the observed spectrum with $E_{i} = 30.5$ meV in Fig.~\ref{Fig_3}(a). 
In model A the constraints $J_{1} = J_{2}$ and $J_{3} = J_{4}$ are imposed. 
The model is based on the crystal structural consideration that the atomic distances for the first (third) and second (fourth) neighbored Ni pairs and the relevant Ni$-$O$-$Ni (Ni$-$O$-$O$-$Ni) bond angles 
are similar, as shown in TABLE.~\ref{Tab_1}. 
Previous Raman scattering study~\cite{PhysRevB.100.144417} used this model. 
In model B no constraint is imposed. 
The calculated spectra using the best fit parameters 
for the models are shown in Figs.~\ref{Fig_3}(b) and~\ref{Fig_3}(c). 
The experiment and calculation of the 
one-dimensional (1D) cuts along the energy axis and those along the momentum axis are shown in Fig.~\ref{Fig_3}(d) and \ref{Fig_3}(e), respectively. 
The best parameters are summarized in Table~\ref{Tab_2} (see appendix A for the detail of the fitting). 
The model B gives a better correlation coefficient $R$ and $\chi ^2$, where $\chi ^{2}=\frac{1}{N} \sum_{i}^{N}{\frac{(S_{i}^{\rm exp}-S_{i}^{\rm sim})^2}{\epsilon_{i}^2}}$ and $\epsilon_{i}$ is the experimental error for $S_{i}^{\rm exp}$. 
\begin{table}
\caption{\label{Tab_2}
Estimated exchange constants in the unit of meV, $\chi ^2$, and correlation coefficient $R$ 
for spin models.}
\begin{ruledtabular}
\begin{tabular}{ccccccc}
&$J_{1}$ &$J_{2}$ &$J_{3}$ &$J_{4}$ &$\chi ^2$&$R$\\
\hline Model A&\multicolumn{2}{c}{3.56}&\multicolumn{2}{c}{0.37}&8.8&0.928\\
Model B&6.05&0.95&0.21&0.19&6.1&0.953\\
\end{tabular}
\end{ruledtabular}
\end{table}

\begin{figure}
\includegraphics[width=\linewidth]{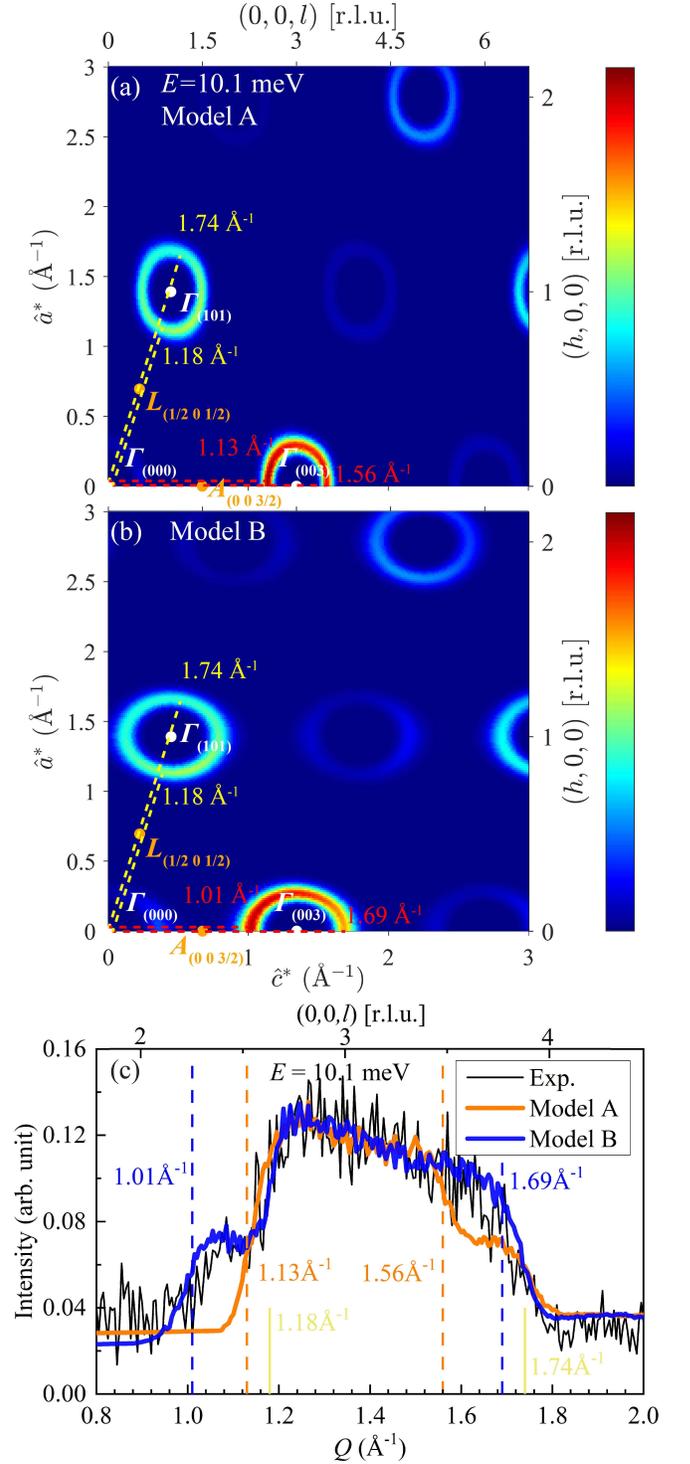}
\caption{\label{Fig_4}(a),(b) 
Simulated spectra of single crystal slice at $E=10.1$ meV for Model A (a) and Model B (b). 
High symmetry points, $\Gamma$, $A$, and $L$ are marked. 
The dash lines denote the radius of momentum sphere first touches and departs the dispersion of ellipses. 
(c) 1D-cut of powder spectrum at $E=10.1$ meV for experiment (black), model A (orange) and model B (blue). 
The vertical dashed orange (blue) lines indicate positions of the first touches and 
departures around $\Gamma_{(003)}$ for model A (B). 
The solid yellow lines indicate those around $\Gamma_{(101)}$ for both models A and B. } 
\end{figure}

\section{Discussion}

The spin wave excitation at 10 K is found to stem from 
$Q \sim 1.4~{\rm \AA}^{-1}$ in the spectrum in Fig.~\ref{Fig_3}(a). 
The momentum approximately corresponds to reciprocal points $(003)$ and $(101)$, which are called $\Gamma _{(003)}$ and $\Gamma _{(101)}$, 
respectively. 
The observed excitation is the superposition of the modes from these $\Gamma$ points. 

In the powder INS spectrum, with the increase of $Q$, the radius of a sphere in the three-dimensional reciprocal space, the INS signal 
appears when the sphere touches a surface of spin wave dispersion with nonzero intensity.
The spectrum, then, loses the intensity when the sphere surface departs 
from the dispersion surface. 
Fig.~\ref{Fig_4}(a) shows the simulated intensity profile of a single crystal for model A 
sliced at $(h0l)$ plane and at $E$ = 10.1 meV. 
The simulated dispersions around $\Gamma$ points are ellipses with major 
axes laying along the $a^{*}$-axis. For Model B in Fig.~\ref{Fig_4}(b), on the contrary, the major 
axes lay along the $c^{*}$-axis. 
These features are caused by the fast and slow velocities of spin waves 
along the $c^{*}$-axis for Model A and Model B, respectively. 
Considering the geometrical relation, the $Q$ positions of contact and departure between the 
momentum sphere and dispersion ellipse around $\Gamma_{(101)}$ are 
almost indistinguishable for Model A and Model B: contact at 
$Q \sim 1.18$ \AA$^{-1}$ and departure at $Q \sim 1.74$ \AA$^{-1}$. 
In contrast around $\Gamma_{(003)}$, the positions are determined by 
the dispersion along the $c^{*}$-axis, leading to a significant difference: 
contacts at $Q \sim 1.13$ \AA$^{-1}$ for model A and 
$Q \sim 1.01$ \AA$^{-1}$ for model B, 
and 
departures at $Q \sim 1.56$ \AA$^{-1}$ for model A and 
$Q \sim 1.69$ \AA$^{-1}$ for model B. 
In the powder-averaged profile in Fig.~\ref{Fig_4}(c), 
the calculation based om model B indicated by the blue curve has a shoulder structure at 
$Q \sim 1.01$ \AA$^{-1}$, ascribed to the different contact $Q$s 
between the dispersions from $\Gamma _{(003)}$ and $\Gamma _{(101)}$. 
In contrast the calculation based on model A indicated by the orange curve does not 
have a shoulder structure at $Q \sim 1.01$ \AA$^{-1}$ because 
the contact $Q$s for the dispersions from $\Gamma _{(003)}$ and $\Gamma _{(101)}$ 
are similar to each other. 
At $Q \sim 1.7$ \AA$^{-1}$ the shoulder is absent for model B 
and present for model A. 
This is understood by considering 
the departure of the $Q$ sphere from the spin dispersion for 
each model. 
The experimental profile indicated by the black curve is reproduced 
better by model B than by model A. 
By combination with better coefficients of $\chi ^2$ and $R$ for model B, 
we can safely conclude that model B is more appropriate than model A.

The spin model in \NISO\ 
turns out to be a coupled two-dimensional (2D) honeycomb lattice 
stacked along the $c$-axis with intraplane interaction $J_1$ of 6.05 meV 
and interplane interaction $J_2$ of 0.95 meV. 
The difference between $J_1$ and $J_2$ is inconsistent with 
Raman scattering study~\cite{PhysRevB.100.144417} 
but consistent with the first principle calculation (GGA+U) for the isostructural compound \NTO~\cite{wu2010theoretical}. 
It should be noted here that $J_{4}$ and $J_{5}$ in \NTO\ in Ref.~\onlinecite{wu2010theoretical} 
correspond to $J_{1}$ and $J_{2}$ in \NISO\ in the present study, respectively. 

The third and fourth neighbor interactions in honeycomb lattice, $J_3$ for Ni(1) and $J_4$ for Ni(2), are both antiferromagnetic,  
leading to competition with antiferromagnetic $J_1$ and $J_2$. 
The magnitudes of  $J_3$ and $J_4$, however, are not large enough to 
induce IC structure. 
Indeed in an isolated classical honeycomb lattice, $J_2/J_1 \gtrsim 0.2$ is required for 
the IC structure~\cite{AsaiPRB16,AsaiPRB17}. 
DM interaction instead of geometrical frustration 
is the major origin of the helical magnetism in \NISO.

DM interactions having polar and chiral components 
are allowed between Ni ions connected by $J_1$ and $J_2$ bonds 
from the crystallographic symmetry. 
Now that $J_1$ and $J_2$ are different by a factor of 6, the magnitudes of the corresponding 
DM interactions are different as well.
The staggered magnitude of the chiral component of the DM interaction induces 
staggered magnetization along 
the propagation direction of the helix at zero magnetic field. 
In the circumstance, the spin soliton lattice can be induced 
in the field applied along the helix axis. 
This agrees with the proposed scenario of the spin soliton lattice 
in Ref.~\onlinecite{PhysRevB.102.054409}. 

Based on the mean field (MF) theory, Weiss temperature is estimated to be 
-171.6 K by using the exchange parameters of model B.
In previous magnetic susceptibility measurements, 
Weiss temperatures were estimated as $-207$ K and $-188$ K 
for magnetic field $H\perp c$ and $H\parallel c$, respectively~\cite{PhysRevB.102.054409}. 
The small discrepancy can be explained by the low dimensionality of the spin system or the weak frustration. 
With a decrease in temperature, the susceptibility of a quasi-2D spin model 
increases more moderately than that in 3D model owing to 
short-range antiferromagnetic spin correlation, 
which leads to a larger estimate of Weiss temperature.

Though the present INS experiments with relax $Q$ resolution do not probe the helimagnetic order, 
we can estimate the chiral component of DM vector along the propagation of the helix 
from the reported propagation vector. 
The calculation of MF energy gives the spin-flop field in the field applied 
along the $c$ axis $H_{c}=18.3$ T (see Appendix B in detail). 
The estimate is consistent with the critical field, 16 - 19 T, previously 
reported in the magnetization measurements~\cite{PhysRevB.102.054409}.

\section{Conclusion}
Inelastic neutron scattering experiment was performed on a proper-screw-type helimagnet 
\NISO\ at zero magnetic field using a polycrystalline sample. 
Well-defined spin-wave excitation was observed. 
The obtained spectrum was carefully compared with the simulated spectra by linear spin-wave theory 
on the basis of two spin models: a three-dimensional spin 
model with the constraint $J_1=J_2$ and 
a coupled honeycomb spin lattice model with $J_1\neq J_2$. 
The latter model with $J_1$ = 6.05 meV and $J_2$ = 0.95 meV well reproduced the observed spectrum. 
The difference between $J_1$ and $J_2$ leads to staggered DM interactions along the polar $c$ axis, 
which is the basis of the soliton lattice scenario in the field applied 
perpendicular to the $c$ axis. 
The critical field of the spin-flop transition in the field applied parallel to the $c$ axis 
estimated by MF calculation was consistent 
with the previous magnetization measurement. 

\begin{acknowledgements}
We are grateful to D. Kawana, T. Asami, and R. Sugiura for supporting us in the neutron scattering experiment at HRC and HER. 
The neutron experiment using HRC spectrometer at the Materials and Life Science Experimental Facility of the J-PARC was performed under a user program (Proposal No. 2021S01). 
The neutron experiment using HER at JRR-3 was carried out by the joint research in the Institute for Solid State Physics, the University of Tokyo (Proposal No. 21403). 
Z. Liu was supported by the Japan Society for the Promotion of Science through the Leading Graduate Schools (MERIT). 
This project was supported by JSPS KAKENHI Grant Numbers 19KK0069, 20K20896 and 21H04441. 
\end{acknowledgements}

\appendix
\section{Fitting Details}
The calculated spin-wave spectrum was modified 
by introducing a spin-wave lifetime and background. 
The fitting function is:
\begin{equation}
S^{\rm sim}(Q,E) = A_{1} {\hat S}^{\rm sim}(Q,E) + A_{2} \cdot Q^{2} + A_{3}.
\label{Eq_A1}
\end{equation}
Here ${\hat S}^{\rm sim}(Q,E)$ is the simulated structure factor convoluted by 
Gaussian function with FWHM = $2\Gamma$, $A_{1}$ is the normalization factor, and 
the second and third terms stand for background.
$\Gamma$ includes both the instrumental energy resolution 
and the energy linewidth generated by spin wave lifetime. 
The best fitting results were obtained by minimizing $\chi ^{2}$. 

The fitting region was selected as 0.8\AA $^{-1}$$<Q<3.55$ \AA $^{-1}$ and 
$3.05~{\rm meV}<E<25.0$ meV 
of the experimental data to exempt incoherent elastic scattering. 
The best fitting result for model A is $J_{1} = J_{2} = 3.56$ meV, $J_{3} = J_{4} = 0.37$ meV, 
$\Gamma = 2.030$ meV and $\chi _{A}^{2}=8.8165$ (correlation coefficient $R_{A} =0.9283$). 
The best result for model B is $J_{1} = 6.05$ meV, $J_{2} = 0.95$ meV,
$J_{3} = 0.21$ meV, $J_{4} = 0.19$ meV,
$\Gamma = 1.622$ meV and $\chi _{B}^{2}=6.0843$ (correlation coefficient $R_{B} =0.9533$). 
The evaluation coefficients $\chi _{A}^{2}>\chi _{B}^{2}$ and $R_{A}<R_{B}$ 
showed a better fitting for model B than for model A.

\section{Spin-Flop Field}
\begin{figure}[htbp]
\centering
\includegraphics[width=\linewidth]{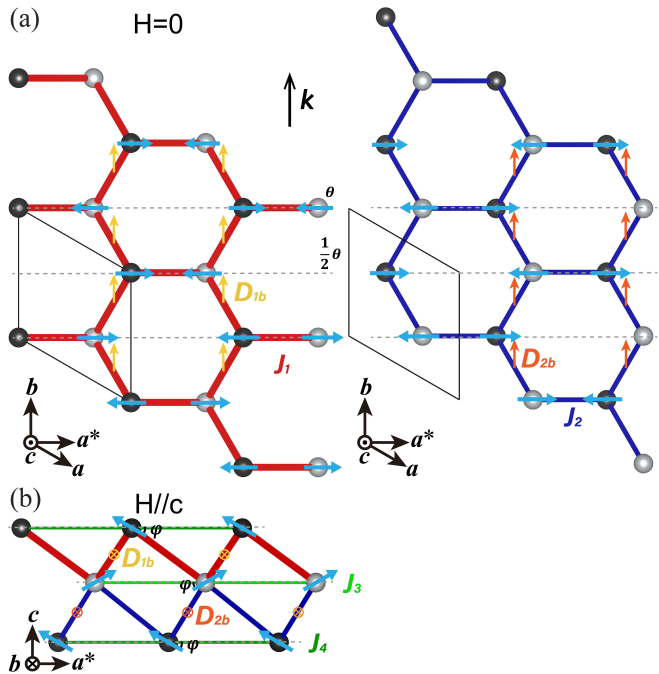}
\caption{\label{Fig_5}(a) Proper-screw-type spin structure at zero field in a $\bm{k}\parallel \bm{b}$ microscopic domain. The light blue arrows represent the component of Ni$^{2+}$ spin moments in the $ab$-plane, and they were scaled up for visualization. 
The spin rotates by $\theta$ with propagating distance $b$. 
$D_{1b}$ and $D_{2b}$ are the chiral components of DM vectors. (b) Canted Antiferromagnetic structure in $\bm{H}\parallel \bm{c}$.} 
\end{figure}
The reported propagation vector $\bm{k}$ is isotropic 
in the $c$-plane in the small-angle soft-X-ray scattering and neutron diffraction experiment~\cite{PhysRevB.102.054409}. 
This means that the compound is in a multi-domain state, which is sensitive to a weak external field such as strain and a magnetic field. 
Here we consider a domain with $\bm{k}\parallel \bm{b}$ 
where the spin moments rotate in the $a^{*}c$-plane. 
We calculated the mean-field energy $E_{\text{helix}}$ at zero field for the spin model shown in 
Fig.~\ref{Fig_5}(a), 
\begin{eqnarray}
E_{\text{helix}}(\theta) &=& -(J_{1}+J_{2})S^{2}(1+2\cos{\frac{1}{2}\theta}) \nonumber \\
&+& (J_{3}+J_{4})S^{2}(\cos{\theta}+2\cos{\frac{1}{2}\theta}) \nonumber \\ &-&2(D_{1b}+D_{2b})S^{2}\sin{\frac{1}{2}\theta}.
\label{Eq_B1}
\end{eqnarray}
Here $\theta$ is defined as the rotation angle with propagating distance $b$.
$D_{1b}$ and $D_{2b}$ are chiral components of DM vectors that 
induce helical structure along the $b$-axis. 
The $c^*$-axis components of the DM vectors which may induce staggered weak canted magnetization along the $b$-axis at zero field were 
neglected because the the canted magnetization has been difficult to be observed. 
We solved the equation $(\frac{\partial E_{\text{helix}}}{\partial \theta})_{\theta = \theta _0}=0$ 
where $\theta _0$ was the rotation angle between the neighboring spins 
under the assumption that the $D/J$ values are common for the bonds $J_1$ and $J_2$. 
We, then, obtained 
$D_{1b}=0.530$ meV and $D_{2b}=0.083$ meV. 

To estimate the spin-flop field in the field applied parallel to the $c$ axis, 
we made an assumption that the helical spin structure was 
not changed by the magnetic field 
below the spin-flop field and the MF energy held. 
In addition, we assumed a canted antiferromagnetic structure (shown in Fig.~\ref{Fig_5}(b)) 
which is a standard structure reported in helimagnets when the field is applied in 
the spin rotation plane~\cite{tokunaga2015magnetic, PhysRevB.69.064114}. 
Then, the MF energy $E_{\text{CAF}}$ is 
\begin{eqnarray}
E_{\text{CAF}}&&(\varphi) = -3(J_{1}+J_{2})S^{2}\cos{2\varphi}+3(J_{3}+J_{4})S^{2}\nonumber \\&&-(D_{1b}-D_{2b})S^{2}\sin{2\varphi}-g\mu_{B}HS\sin{\varphi},
\label{Eq_B2}
\end{eqnarray} 
where $\varphi$ is the canted angle. 
We solved the equations $\frac{\partial E_{\text{CAF}}}{\partial \varphi}=0$ and 
$E_{\text{helix}}=E_{\text{CAF}}$, 
and $\varphi =0.039$ rad and critical field $H_{c}=18.3$ T were obtained.
Here $g=2.26$ was used for Ni ions according to magnetization experiments 
in Ref.~\cite{doi:10.1021/cm304095s}.

\bibliography{References221208.bib}

\end{document}